# DSRS: Estimation and Forecasting of Journal Influence in the Science and Technology Domain via a Lightweight Quantitative Approach

Snehanshu Saha[1], Neelam Jangid[2], Archana Mathur[3] Anand M N[4]
[1,2,3]Department of Computer Science and Engineering PESIT South Campus Bangalore-560100
[4]BITS Hyderabad
E-mails: [1]snehanshusaha@pes.edu  [2]neelu.jangid88@gmail.com  [3]archanamathur@pes.edu
[4]anand@hyderabad.bits-pilani.ac.in

**Abstract**: The evaluation of journals based on their influence is of interest for numerous reasons. Various methods of computing a score have been proposed for measuring the scientific influence of scholarly journals. Typically the computation of any of these scores involves compiling the citation information pertaining to the journal under consideration. This involves significant overhead since the article citation information of not only the journal under consideration but also that of other journals for the recent few years need to be stored. Our work is motivated by the idea of developing a computationally lightweight approach that does not require any data storage, yet yields a score which is useful for measuring the importance of journals. In this paper, a regression analysis based method is proposed to calculate Journal Influence Score. Proposed model is validated using historical data from the SCImago portal. The results show that the error is small between rankings obtained using the proposed method and the SCImago Journal Rank, thus proving that the proposed approach is a feasible and effective method of calculating scientific impact of journals.

**Keywords:** Journal Influence Score (JIS), Downselection with Regression and Significance scheme (DSRS), Multiple Linear Regression (MLR), Clustering, Significance test, Internationality, Principal representative features.

1. Introduction and Background

Librarians and information scientists have been evaluating journals with regard to importance and popularity for the last 75 years. The Journal Impact Factor proposed by Eugene Garfield [3] the founder and Chairman Emeritus, ISI (which later became Thomson Reuters) was a milestone in this regard. The advent of the Thomson Reuters citation indexes made it possible to do computer-compiled statistical reports in terms of citation frequency. Thomson Reuters is credited with the invention of the journal "impact factor" in the '60s and began to publish Journal Citation Reports in 1975 as part of the *SCI* and the Social Sciences Citation Index [2]. Scimago journal rank (SJR) is a relatively recent method of ranking the journals in Science and technology. Scopus and ISI are two other well established schemes.

The evaluation of journals based on their influence is of interest for numerous reasons. Increasingly both academic and research institutions are using publication information for evaluation of faculty/employees. It also has significant policy implications both at institution level as well as at country level. For instance, the Department of Science and Technology (DST), Government of India commissioned an Evidence based a study through Thomson Reuters. This focused on the analysis of changing trends in publication habits of the Indian scientific community the main motivation being to inform and assist policy bodies and funding decisions in the country. As the volume of such journals in the Science and technology space is increasing in astronomical proportions, as evident from the Department of Science and Technology (DST) report [**15**] also note that this relates to India alone), some clarity is needed for the authors who wish to publish in such journals as well as the institutions/employers that judge such publications and use those for appraisal or funding decisions.

While there exists methodologies for journal evaluation as mentioned earlier, these are insufficient considering the astronomical growth in the number of journals. The SJR rank for instance is just a numerical score, and therefore can't directly be used as threshold. SJR is also dependent heavily on five years' data and addresses a limited list of journals. Scopus and ISI are excellent datasets of evaluating journals and the quality of articles published as long as the total number of journals published around the world has a one to one correspondence with the number of publications listed in either of these two abstracting/indexing services. Unfortunately, these two services combined cater to less than 1% of the volume of journals published.

The number of journals in many fields has grown significantly over the last few years. India is a case in point, illustrating almost exponential growth in journal proliferation. Particularly in the year 2009, there was remarkable increase of 25% in scientific publications than the previous year. There's a lot of data available for Indian journals on quality metrics. Although significant bibliometric information has been documented [14], there are limited numbers of studies dealing with analysis. One such work is an exploratory analysis of Indian science and technology publication output conducted by Buchandiran [2]. Apart from analysis of trends, the author also states that when an Indian author writes a qualitative scientific paper, he/she likes to publish the paper in an international reviewed journal. The simple reason is that barring a very few, most of the Indian Science and Technology journals lack perceived quality and the reception to them at the international level is poor. The number of Indian journals covered in the international databases such as *ISI* and *Scopus* is very limited.

A classical way of measuring Journal influence/reputation is the impact factor of the journal [3]. The impact factor is usually calculated over a period based on the number of citations in the current year to articles in the journal during the previous years. For instance, the two year impact factor is calculated for year **n** as below.

Let $A$ = the number of times all items published in a journal in years (**n-1**) and (**n-2**) that were cited by indexed publications during year **n**.
Let $B$ = the total number of "citable items" published by that journal in years (**n-1**) and (**n-2**). ("Citable items" for this calculation are usually articles, reviews, proceedings, or notes; not editorials or letters to the editor).
Then the impact factor for year **n** = $A/B$.

Despite some of the criticisms, the Journal Impact Factor is widely used as a measure of the scientific influence of a journal. Most of the variants of the Journal Impact Factor require the citation data for the preceding few years from all the indexed journals. Also note that the impact factor for year **n** is only available in the next year (**n+1**) because all the citations to previous years need to be available. Thus the computation of the impact factor of a journal involves significant overhead.

In this paper, a new approach for calculating Journal Influence Score (JIS) is proposed. This paper is motivated by the idea of computing JIS without the overhead of citation data storage but at the same time yielding reasonable accuracy. We model it as a regression problem in which the individual weights corresponding to each of the input variables are computed from historical data, these weights reflect the relative influence the individual variables may have in the calculation of the influence score. These weights can then be used directly to compute the score of a journal without using the prestige of any other journal. The main advantage is a computationally lightweight scheme that does not require any data storage [8] .We conducted experiments on publicly available data to validate our approach.

The remainder of the paper is organized as follows. The next section, 2 elucidates the overview of the model and the algorithmic scheme. Section also highlights key statistical terminologies used throughout the paper. Section 3 is a walk-through of DSRS, a mixture of MLR, subset selection and significance tests. This section leads us to computing JIS, which is then used for the next task. Section 4 presents the results of the DSRS model, and the online tool for computing JIS. Once JIS is computed, section 5, then elaborates the clustering process of the journals based on the JIS score. Section 6, summarizes the model, outcome and experimental efficacies. The final section 7, throws a lot of open questions based on the JIS score and a very important problem in scientometrics, that is, internationality of journals. Appendix contains information about the detailed subset selection method and code for the toolkit.

## 2. The Model: Overview and Algorithmic Flow:

This section describes the details of our approach. We use a linear regression model where the response variable is the Journal Influence Score. The years under consideration were 2011, 2012. The input parameters (predictor variables) include the Quarter, H-Index, Total Docs 2012, Total Docs 3yrs, Total Cites 3yrs, Citable Docs 3yrs, Ref/Doc, Cites / Doc. (4years), Cites/Doc (3yrs), Cites / Doc. (2years) and Total Ref, Cited Docs, Uncited Docs, %International Collaboration (cf. Table 2)

Along with the 13 parameters picked up from SCImago Journal and Country Rank, Quarter was also considered as one of the input variables. Intuitively, any journal to be evaluated in the first Quarter of the year has more probability of having greater influence, considering the number of publications is mostly limited. Hence the "quarter" of publication should be statistically significant. The probability of influence in our sample data validates the use of quarters in our model.

$$\text{Quarter(Probability of Influence)} = \frac{Q_1}{\sum Q_1} \quad \text{Where i = 1, ... ,4} \tag{1}$$

Starting with the initial set of input parameters, a two-phase approach was employed to obtain a more compact set of transformed variables. In the first phase, the number of variables was reduced using cross-correlation & MLR, and a down selected set of input variables was obtained. In the second phase, DSRS was applied on this reduced which generates a result that shows what percentage of the variability is explained by the given dataset. Pair wise correlation and retaining only those pairs with minimal correlation helps in reduction of input factors while maintaining the reasonable accuracy. The final model was a MLR model on the principal representative features retained after the second phase.

The SCImago Journal & Country Rank (SJR) [21] is a portal that includes the journals and country scientific indicators developed from the information contained in the Scopus database. These indicators can be used to assess and analyze scientific domains. Our source data for this study were SCImago Journal and Country Rank's portal which contained journals in Elsevier's Scopus. [22]

### 2.1 Algorithm

Step1. Import data from web (www.scimagojr.com)

Step2. Find correlation of all factors with SJR

$$R_{x1x2} = \frac{\sum x_{1i} x_{2i} - n\bar{x}_1 \bar{x}_2}{[\sum x_{1i}^2 - n\bar{x}_1^2]^{\frac{1}{2}} [x_{21}^2 - n\bar{x}_2^2]^{\frac{1}{2}}} \tag{2}$$

Step3. Find Model equation of type

$$y = bx + e \tag{3}$$

Step4. Find parameter b

$$b = (X^T X)^{-1}(X^T y) \tag{4}$$

Step5. Derive Multiple Linear Regression equation to establish relationship between input factors and journal ranking output

Step6. Extract p-values and correlation coefficient values from Multiple Linear Regression equation

Step7. If input variables with P-value > 0.05 & Correlation Coefficient < 0.4, then remove parameter.

Step8: Repeat Step 7 for all Input factors

Step9: Repeat Step 2 – 8 till all parameters has P-value < 0.05 & Correlation Coefficient > 0.4

Step10. Compute the mean and standard deviations of the variables.

$$\bar{x}_s = \frac{1}{n}\sum_{i=1}^{n} x_{si} \qquad s_x^2 = \frac{1}{n-1}\sum_{i=1}^{n}(x_{si}-\bar{x}_s)^2 \qquad (5)$$

Step11. Normalize the variables to zero mean and unit standard deviation.

$$x'_s = \frac{x_s - \bar{x}_s}{S_x} \qquad (6)$$

Step 12. Compute the correlation among the variables:

$$R_{x_s,x_r} = \frac{\frac{1}{n}\sum_{i=1}^{n}(x_{si}-\bar{x}_s)(x_{ri}-\bar{x}_r)}{S_{xr}S_{xs}} \qquad (7)$$

Step13. Prepare the correlation matrix:

Step14. Compute the Eigen values & Eigen vectors of the correlation matrix.

Step15. Obtain principal representative features by multiplying the eigenvectors by the normalized vectors

Step16. Compute the values of the principal representative features.

Step17. Compute the sum (the sum must be zero) and sum of squares of the principal representative features. The sum of squares gives the percentage of variation explained.

Step18. Apply regression on the Principal factors to compute JIS.

Step19. Calculate the quartile match
    19.1: Check the samples in each quarter.

    19.2: Compare these samples with same number of samples from the results of our model for each quartile.

    19.3: Calculate the percentage of match.

Table 1 below summarizes a comparison between the existing method and the proposed method in terms of number of input variable, procedure used; time complexities, database size, and algorithmic approaches. Table 2 shows initial 13 indicators used from SCImago Journal and Country Rank's portal. Out of these 13 input variables, Cites / Doc. (4years), Cites / Doc. (3 years) are not considered for evaluation with an intention of working with a smaller dataset since the proposed method considers data of two years only. This leaves us 11 input/predictor variables to work with (Table 3). Further, as the data for Cited Docs, Uncited Docs and International Collaboration is sparse and unavailable for most of the journals, these parameters are removed from evaluation. So the model works on remaining 8 variables and "Quarter" being the 9$^{th}$ parameter (Table 4) at the first iteration step.

| Basis of Comparison | Existing Method | Proposed Method |
|---|---|---|
| Number of input variables | 13 | 5 |
| Procedure | Iterative | Weighted |
| Expected Time Complexity | More | Less |
| Database Size | Huge | Insignificant |
| Historical Data | 5 Years | 2 Years |
| Algorithm Used | Google Page Rank Algorithm | Weight based on regression |

**Table 1: Comparison Between Existing & Proposed System**

SCImago JR database publishes data for 13 input variables (cf. Table 2). The "Ab initio" model proposed in the paper uses the data and eventually the DSRS model evolves which requires ONLY five input variables explaining the transition/transformation from 13 to 5.

| Year | 2008 | 2009 | 2010 | 2011 | 2012 |
|---|---|---|---|---|---|
| Total Documents | 34 | 26 | 20 | 21 | 23 |
| Total Docs. (3years) | 25 | 59 | 85 | 80 | 67 |
| Total References | 662 | 529 | 515 | 776 | 846 |
| Total Cites (3years) | 15 | 37 | 51 | 74 | 92 |
| H-Index | 4 | 2 | 1 | 14 | 30 |
| Citable Docs. (3years) | 25 | 59 | 85 | 80 | 67 |
| Cites / Doc. (4years) | 0,60 | 0,63 | 0,60 | 0,99 | 0,97 |
| Cites / Doc. (3years) | 0,60 | 0,63 | 0,60 | 0,93 | 1,37 |
| Cites / Doc. (2years) | 0,60 | 0,63 | 0,33 | 1,30 | 1,68 |
| References / Doc. | 19,47 | 20,35 | 25,75 | 36,95 | 36,78 |
| Cited Docs. | 10 | - | - | -- | -- |
| Uncited Docs. | -- | 1 | -- | -- | 1 |
| %International Collaboration | 6,90 | -- | -- | --- | --- |

**Table 2: Sample data from SJR**

| Year | 2008 | 2009 | 2010 | 2011 | 2012 |
|---|---|---|---|---|---|
| Total Documents | 34 | 26 | 20 | 21 | 23 |
| Total Docs. (3years) | 25 | 59 | 85 | 80 | 67 |
| Total References | 662 | 529 | 515 | 776 | 846 |
| Total Cites (3years) | 15 | 37 | 51 | 74 | 92 |
| H-Index | 4 | 2 | 1 | 14 | 30 |
| Citable Docs. (3years) | 25 | 59 | 85 | 80 | 67 |
| Cites / Doc. (2years) | 0,60 | 0,63 | 0,33 | 1,30 | 1,68 |
| References / Doc. | 19,47 | 20,35 | 25,75 | 36,95 | 36,78 |
| Cited Docs. | 10 | - | - | -- | -- |
| Uncited Docs. | -- | 1 | -- | -- | 1 |
| %International Collaboration | 6,90 | -- | -- | --- | --- |

**Table 3: Selected 11 Parameters**

**NOTE:** It has been determined that regression model applied after the Downselected optimal feature selection, proposed here accounts for the fact that the principal representative features are linearly independent of each other in the domain space. K-means Clustering is implemented to achieve classification between "National" and "International" journals, a metric that libraries and academic institutions may use for measuring impact of scientific work.

> **Key Terms:**
>
> **Null Hypothesis considered for significance testing:** Input factors are not able to describe the Output.
>
> **F-Test:** This test is used to check the hypothesis that the proposed model fits the data well. The model is significant if MSR/MSE value is greater than F[k,n-k-1] which is taken from F-distribution table.
>
> **P-value:** P-value is the probability of obtaining a test statistic at least as extreme as the one that was actually observed, assuming that the null hypothesis is true. Hypothesis is rejected if the value is less than 0.05 for 95% confidence interval.
>
> **Hypothesis Testing:** Hypothesis Testing is a test of significance and used to support a decision statistically. Hypothesis testing is divided into three steps.
>
> **Step 1:** Define Null and alternative hypotheses
> <u>Null Hypothesis</u> (Ho): The null hypothesis is a claim of no difference or no impact
> <u>Alternative Hypothesis</u> (Ha): Opposite of null hypothesis, a claim of difference or impact
>
> **Step 2:** Calculate test statistics and p-value
> * P(Probability) Value
>     - Statistical measure for the strength of Ho
>     - Probability that it will be wrong to select the Alternate Hypothesis
>     - Higher the p value, the more evidence we have to support Ho
>     - P-value less than 0.05 means we reject the null hypothesis.
>
> **Step 3:** Take statistical significance decision : By convention if p >.05, Accept H0
>
> 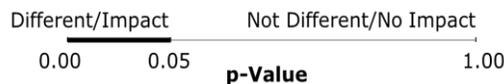
>
> If p≤.05, Reject Ho and Accept Ha
> **F-value**: It is used for significance testing for ANOVA (Analysis of Variance)

## 3. Description of Model: Multiple Linear Regression Model:

Linear regression is an approach for modeling the relationship between a dependent variable y and one or more explanatory variables denoted by x. When one explanatory variable is used, the model is called simple linear regression. When more than one explanatory variable are used to evaluate the dependent variable, the model is called multiple linear regression model.

Applying multiple linear equation model to predict a response variable y as a function of 9(initially 13, reduced to 9) predictor variables $x_1, x_2, x_3 \ldots x_9$ takes the following form:

$$y = b_0 + b_1 x_1 + b_2 x_2 \ldots \ldots b_9 x_9 + e \tag{8}$$

Here, $\{b_0, b_1 \ldots b_9\}$ are 10 fixed parameters and e is the error term.

Given a sample $\{(x_{11}, x_{21}, x_{31}, \ldots x_{91}, y_1), \ldots (x_{1n}, x_{2n}, x_{3n}, \ldots x_{9n}, y_n)\}$ of n observations the model consist of following n equations

$$y_1 = b_0 + b_1 x_{11} + b_2 x_{21} \ldots \ldots b_9 x_{91} + e_1 \tag{9}$$
$$y_2 = b_0 + b_1 x_{12} + b_2 x_{22} \ldots \ldots b_9 x_{92} + e_2 \tag{10}$$
$$y_3 = b_0 + b_1 x_{13} + b_2 x_{23} \ldots \ldots b_9 x_{93} + e_3 \tag{11}$$
$$\vdots$$
$$y_n = b_0 + b_1 x_{1n} + b_2 x_{2n} \ldots \ldots b_9 x_{9n} + e_n \tag{12}$$

So, we have

$$\begin{bmatrix} y_1 \\ \vdots \\ y_n \end{bmatrix} = \begin{bmatrix} 1 & x_{11} & \cdots & x_{k1} \\ \vdots & \vdots & & \vdots \\ 1 & x_{1n} & \cdots & x_{kn} \end{bmatrix} \begin{bmatrix} b_0 \\ \vdots \\ b_k \end{bmatrix} + \begin{bmatrix} e_1 \\ \vdots \\ e_n \end{bmatrix}$$

Where k = {1...9}  (13)

Or in matrix notation: **y** = **Xb** + **e**

where
b = A column vector with 10 elements = {$b_0, b_1, ..., b_9$}
y = A column vector of *n* observed values of y = {$y_1, ..., y_n$}
X = An *n* row by *10* column matrix whose (i,j + 1)th element $X_{i,j+1}$ = 1 if j = 0 else $x_{ij}$

Parameter estimation: **b** = $(\mathbf{X^T X})^{-1}(\mathbf{X^T y})$  (14)

Allocation of variation:

$$SSY = \sum_{i=1}^{n} y_i^2 \qquad SS0 = n\bar{y}^2$$
$$SST = SSY - SS0 \qquad SSE = \{\mathbf{y^T y} - \mathbf{b^T X^T y}\}$$
$$SSR = SST - SSE$$

(15)

Where  SSY – sum of squares of y,    SST – total sum of squares
SS0 - sum of squares of $\bar{y}$,    SSE – sum of squared errors
SSR – sum of squares given by regression

Coefficient of determination:

$$R^2 = \frac{SSR}{SST} = \frac{SST - SSE}{SST}$$

(16)

Coefficient of multiple correlation

$$R = \sqrt{\frac{SSR}{SST}}$$

(17)

Analysis of variance:

$$MSR = \frac{SSR}{k}; \quad MSE = \frac{SSE}{n - k - 1}$$

(18)

Where MSR- Mean square due to regression
MSE- Mean square error

Standard deviation of errors:
$$s_e = \sqrt{MSE}$$
(19)

Standard deviation of parameters:

$$s_{b_j} = s_e \sqrt{C_{jj}}$$

(20)

Where $C_{jj}$ is the *j*th diagonal term of **C** = $(\mathbf{XTX})^{-1}$

New predicted influence score, taking estimated parameters as weights to factors.
Correlations among predictors:

$$R_{x_1 x_2} = \frac{\sum x_{1i} x_{2i} - n \bar{x}_1 \bar{x}_2}{\left[\sum x_{1i}^2 - n \bar{x}_1^2\right]^{1/2} \left[\sum x_{2i}^2 - n \bar{x}_2^2\right]^{1/2}}$$

(21)

This regression model helps us to find the correlated factors so that they can be eliminated, resulting in minimum the number of features. The process helps in classification into the two categories. The output of this model will be the correlation matrix using which are computed using the most correlated factors.

**3.1 Analysis of Multiple Linear regression model:**

**Analysis Phase-I:** After removing 3 parameters (Cited Docs, Uncited Docs, International Collaboration ) from Table 3 (11 parameters) and adding "Quarter" to the table as additional parameter, the final 9 parameters (reflected in Table 4) are fed to analyze the Correlation and Regression statistics.

| Factor | P-value | Correlation Coefficient | Optimization Decision |
|---|---|---|---|
| Quarter | 3.67E-09 | -0.76553 | Yes |
| H index | 0.197845 | 0.691082 | Yes |
| Total Docs. (2012) | 8.84E-05 | 0.155572 | Yes |
| Total Docs.(3years) | 0.902654 | 0.370305 | Yes |
| Total Refs. | 0.008454 | 0.395752 | Yes |
| Total Cites (3years) | 2.32E-07 | 0.554409 | Yes |
| Citable Docs. (3years) | 0.607635 | 0.370423 | Yes |
| Cites / Doc. (2years) | 2.48E-14 | 0.848587 | Yes |
| Ref. / Doc. | 0.82342 | 0.170068 | No |

**Table 4: Analysis phase-I optimization decision table**

Inference:

- The value of coefficient of determination, $R^2$ is 0.7986 i.e.79.86% variation in Journal Influence factor is explained by this regression.
- Significance F value is 8.18E-70 i.e. less than 0.05, which means it passes F-test.
- P-value for Ref/Docs is greater than 0.05 and has a weak correlation with SJR Score. So, Ref/Doc. is removed from further analysis.

**Analysis Phase-II:** Result of Correlation and Regression after removing Ref/Doc is as shown in below table:

| Factor | P-value | Correlation Coefficient | Optimization Decision |
|---|---|---|---|
| Quarter | 3.29E-09 | -0.76553 | Yes |
| H index | 0.1994 | 0.691082 | Yes |
| Total Docs. (2012) | 8.73E-05 | 0.155572 | Yes |
| Total Docs. (3years) | 0.884709 | 0.370305 | No |
| Total Refs. | 0.007956 | 0.395752 | Yes |
| Total Cites (3years) | 2.16E-07 | 0.554409 | Yes |
| Citable Docs. (3years) | 0.61594 | 0.370423 | Yes |
| Cites / Doc. (2years) | 5.81E-15 | 0.848587 | Yes |

**Table 5: Analysis phase-II optimization decision**

**Inference:**

- $R^2$ is 0.7986 i.e.79.86% variation in JIF is explained by this regression, which is similar to the previous $R^2$ (before removing Ref/Doc).
- Significance F value is less than 0.05 i.e. it passes the F-test.
- P-value for Total Docs (3Years) is greater than 0.05 and has a weak correlation with SJR Score. So, Total Docs (3years) is removed from further analysis.

**Analysis Phase-III:**

Regression and Correlation is applied on the data after removing Total Docs (3years). Result is as shown in below table:

| Factor | P-value | Correlation Coefficient | Optimization Decision |
|---|---|---|---|
| Quarter | 1.94E-09 | -0.76553 | Yes |
| H index | 0.199849 | 0.691082 | Yes |
| Total Docs. (2012) | 8.45E-05 | 0.155572 | Yes |
| Total Refs. | 0.007462 | 0.395752 | Yes |
| Total Cites (3years) | 2E-07 | 0.554409 | Yes |
| Citable Docs. (3years) | 0.004463 | 0.370423 | Yes |
| Cites / Doc. (2years) | 3.11E-15 | 0.848587 | Yes |

**Table 6** Analysis phase-III optimization decision table

**DSRS:** Down selection based on pair wise correlation of the set of input variables obtained in previous step. (DOMINATING OPTIMAL FEATURE SUBSET SELECTION)

The downs selected set of variables computed in previous step for multiple journals was used to compute the overall variance from the covariance matrix. We computed pair wise correlations and identified a smaller set of variables such that the correlation between any two variables in this set was small. They can then be used to compute the percentage of variability accounted for individually as shown in table 7. This reduced the number further to only five input variables. The $R^2$ value was very similar to when 9 input variables were considered. We did not do a Principal Component Analysis (PCA) since we were interested in down-selection of features. While in PCA the principal components are orthogonal to each other by design and it provides an elegant way of dimensionality reduction based on percentage variability explained, one problem is interpretation of the transformed variables with respect to the original input variables.

| | |
|---|---|
| Quarter | 62.7952% |
| H-index | 21.3912% |
| Total Docs. (2012) | 8.4489% |
| Total References | 3.539% |
| Cites/Doc(2years) | 1.3498% |
| Citable Docs. (3years) | 1.2379% |
| Total Cites(3years) | 1.2379% |

**Table 7: Percentage of total variability accounted for by individual input variables as calculated in previous step.**

As further validation a regression was run where the small set of five input variables selected as described.

**SUMMARY OUTPUT**

*Regression Statistics*

| | |
|---|---|
| **Multiple R** | 0.8774 |
| **R Square** | 0.76983 |
| **Adjusted R Square** | 0.764575 |
| **Standard Error** | 0.323211 |
| **Observations** | 225 |

**Feature subset selection sequence**

13 (Table 2)→11(Table 3)→ 9 (Table 4)→8(Table 5)→ 7 (Table 6)→5(Table 7)

**ANOVA**

| | Df | SS | MS | F | Significance F |
|---|---|---|---|---|---|
| **Regression** | 5 | 76.51797 | 15.30359 | 146.4943 | 8.28932E-68 |
| **Residual** | 219 | 22.87793 | 0.104465 | | |
| **Total** | 224 | 99.39591 | | | |

| | Coefficients | Standard Error | t Stat | P-value | Lower 95% | Upper 95% |
|---|---|---|---|---|---|---|
| **Intercept** | 0.513322 | 0.110512 | 4.644933 | 5.87E-06 | 0.295518325 | 0.731126 |
| **Quarter** | -0.14076 | 0.030902 | -4.55526 | 8.69E-06 | -0.201667404 | -0.07986 |
| **H index** | 0.004716 | 0.001247 | 3.781766 | 0.000201 | 0.002258486 | 0.007174 |
| **Total Docs. (2012)** | 0.000131 | 0.000121 | 1.084002 | 0.279556 | -0.000107049 | 0.000369 |
| **Total Refs.** | -8.3E-06 | 7.75E-06 | -1.0703 | 0.285661 | -2.35727E-05 | 6.98E-06 |
| **Cites/Doc. (2years)** | 0.301404 | 0.030123 | 10.00582 | 1.19E-19 | 0.242036313 | 0.360772 |

**Table 8: Regression Statistics for analysis phase-III**

**Inference:**

- $R^2$ is 0.7698 i.e. 76.98% variation in Journal Influence score is explained by this regression.
- Significance F value is less than 0.05, which means it passes F-test.

**4. Results:**

**Regression equation:**

The fitting model assumes the form:
***Journal Influence Score = 0.513322- (0.14076 \*Quarter) + (0.004716 \* H index) + (0.000131 \* Total Documents For Current Year) - (8.3E-06 \* Total References) + (0.301404 \*Cites/Doc. in previous 2 years)*** (22)

     The above equation shows that the journal base score is 0.513322, which explains the initial score of each journal, affected further by the factor values of that journal. The other values in the equation signify the weight to

their corresponding factors. Using this equation a Journal Influence score calculator is built which shows an influence score based on the five parameter values of a journal. JIS calculator is available at [23] A video tutorial is available online [25]. Screen shots of the tool are shown in Appendix B.

A comparison has been done to analyze Match % among Journal Ranking using SJR Model, 5-parameter regression model and 9-parameter regression model. Results are as follows:

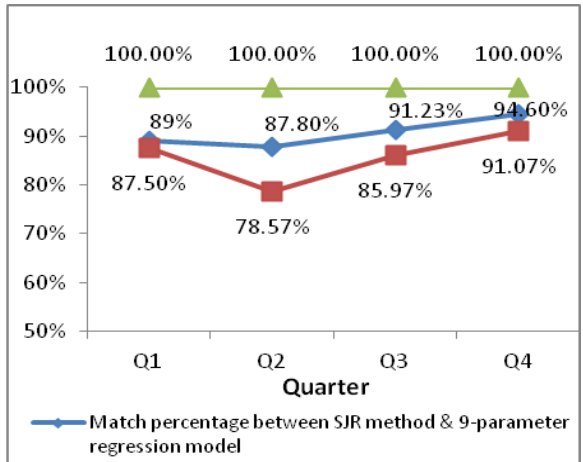 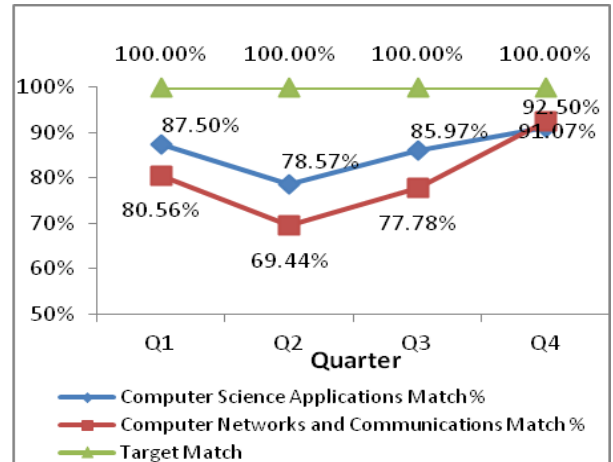

**Fig. 1: Accuracy Test between SJR model & proposed model**

**Fig. 2 Accuracy Test between Computer Science Applications & Computer Networks and Communications**

Computed errors in the proposed model reflect reasonable level of accuracy in all the quarters possible. This comparison is made with the SJR data. Hence, the model performs well in all possible cases and quarters. Another validation has been performed on a different subject category (Computer Networks and Communications) to analyze match % and the results are as follows:

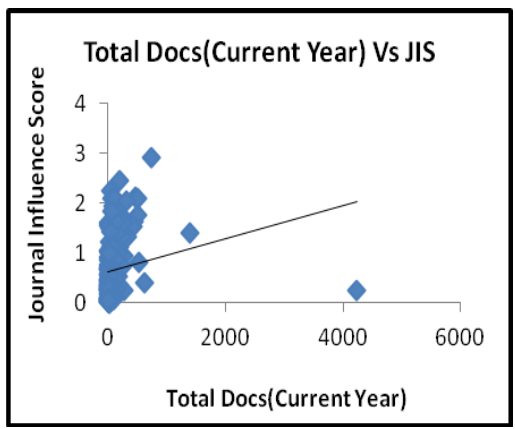 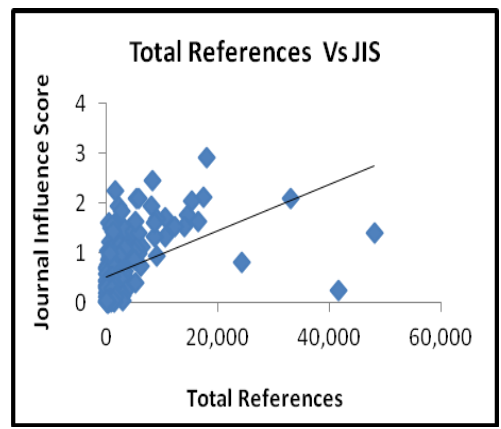

**Fig. 3 Scatter Plot between Total Docs(current year) and JIS**   **Fig. 4 Scatter Plot between Total References and JIS**

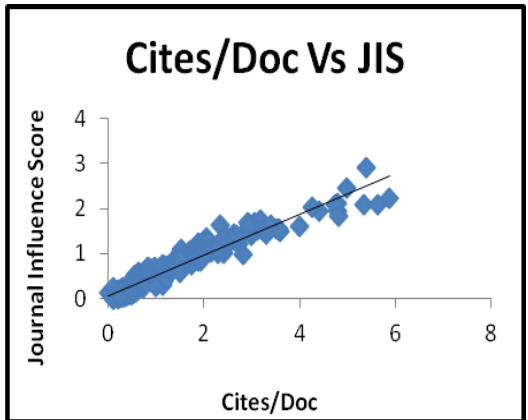 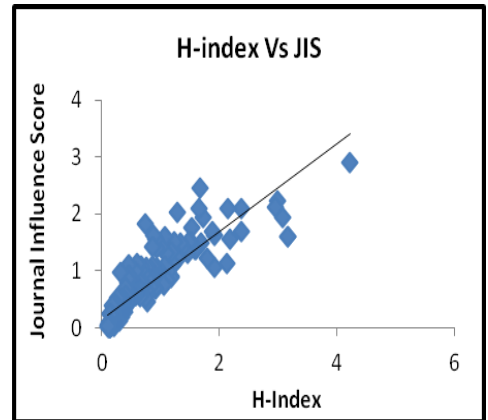

**Fig. 5 Scatter Plot between Cites/Documents and JIS**  **Fig 6 Scatter Plot between H-index and JIS**

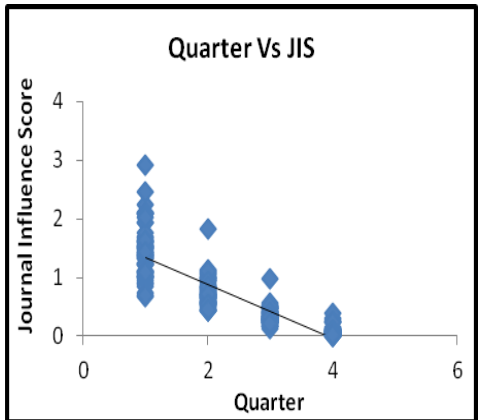

**Fig 7. Scatter Plot between Quarter and JIS**

### 5. <u>Classification Process:</u>

After the influence score for all the journals in the sample set or a new upcoming journal is calculated, the process of classification starts where the intention is to classify the journal into one of the categories, which are National Journal and International Journal. This process is based on the value of the influence score. The higher the influence score, the more the journal is valued and accepted throughout. This is hence used as the means for classification.

The challenge here is to define a boundary that separates the two given classes based on which the class is to be assigned to any journal. This is done by making use of the K-Means Clustering algorithm. Considering the sample set this problem deals with is without the class instances and the training of the system or the machine is unsupervised, the clustering algorithm to be used also has to be unsupervised. Hence, Unsupervised K-Means Clustering Algorithm is put to use.

The samples are clustered and then rearranged iteratively until we have an instance where the change in the cluster means for both the classes is minimal after which the system attains stability. Another point to be kept in mind during this iteration step is that the system should not keep looking for such changes throughout and turn into an infinite loop. This would lead to system deadlock and the system would in-turn fail to perform in the desired manner. This highlights the requirement of an upper limit to the number of iterations performed for this process. Hence maintaining the system stability and also performing well to give the desired results.

### 5.1 K-Means Clustering:

Once the process of reducing the number of factors we take into consideration for the final influence score computation using the DSRS model, we move forward to the part where we differentiate the entities in the structure into National and International categories. We achieve this by making use of the Unsupervised K-Means Clustering method.

### 5.2 The Method:

The samples are clustered into two separate groups in this case by taking two distinct mean points for the clusters initially. These clusters continually keep changing after every iteration based on the evaluation of the change observed in the cluster mean after all the samples in the sample are collected into them. A sample goes into that cluster which lies closer to the position of the sample. Once the clustering is done, we recomputed the means for both the clusters. These steps are put into iteration until we either see negligible change in both the cluster means and the iteration step exceeds a certain limit. This is how we handle clustering of the complete sample set.

### 5.3 Clustering Algorithm:

Step 1: Calculate the influence score of all the journals in the sample set.

Step 2: Select two distinct cluster means arbitrarily.

Step 3: Initialize the variables (Iteration no =0, maxiterations =100, changed = 1)

Step 4: Loop until both the conditions are satisfied While(changed ==1 & iteration no. < maxiterations)

Step 4.1 Increment iteration no, make changed=0

Step 4.2 For all samples in the dataset, classify all into the class with the nearest cluster mean.

Step 4.3 Initialize variables to 0 ($ele_0 = 0$, $ele_{11} = 0$, $sum_0 = 0$, $sum_1 = 0$)

Step 4.4 Re-compute cluster means

For all samples in the dataset

  If ( class == 0)

    Add influence score to $sum_0$, increment $ele_0$

  else

    Add influence score to $sum_1$, increment $ele_1$

    $new_0 = sum_0 / ele_0$;

    $new_1 = sum_1 / ele_1$;

Step 4.5 Check for any significant change in the Cluster Means

    (The threshold value to define a change as stable is anything less than 0.01 [square of 0.1])

$$If\ (((new_0 - u_0))^2 > 0.01\ or\ (new_1 - u_1)^2 > 0.01 \tag{13}$$

    changed = 1

Step 4.6 Store new values of the cluster means      $u_0 = new_0\ \&\ u_1 = new_1$

Step 5: Once the loop terminates, $u_0$ and $u_1$ are the final Cluster Means.

After the execution of the clustering process, the cluster means can be used further for any upcoming new journal entry to make the classification process more and more simple and reducing the complexity. Any new journal just undergoes the influence score calculation process proceeded by a check on this score, if the score lies closer to the cluster mean to the National category, then the journal is termed to be of National Standards, else the journal attains International Standard as the Influence Score lies closer to the cluster that corresponds to the cluster formed of samples of International standard. Hence, classification via the Clustering mechanism is easily achieved and it still maintains the accuracy measures.

The entire algorithm is represented in a flowchart shown below (Fig 8).

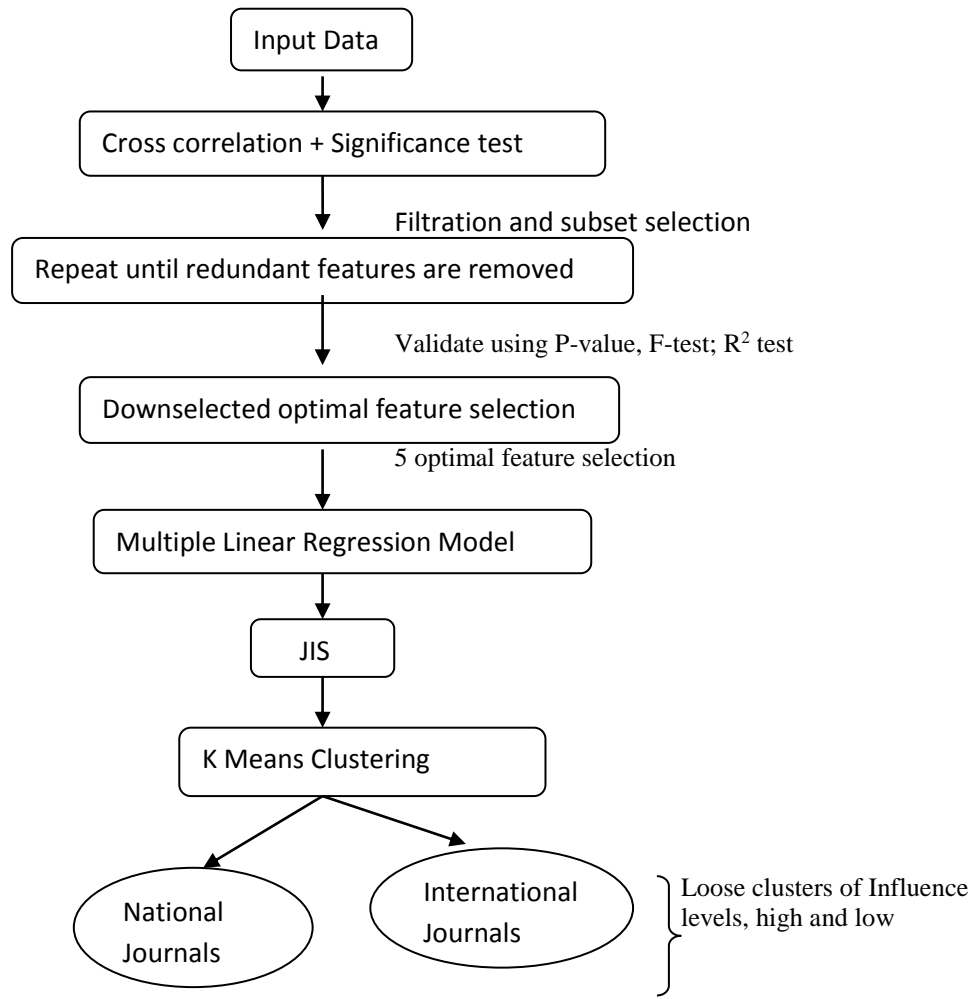

**Fig 8 Flowchart of the Evaluation Model (DSRS)**

**5.5 Outputs and Graphical Plots:**

Here are the plots generated for 1084 samples given for the computer science category, taken from the SJR (SCIMago Journal Ranking) database globally available. The plots show the scatter of the samples and also show the classes in different colors so as to understand how these factors influence the selection process for any sample and to what extent this effect is seen.

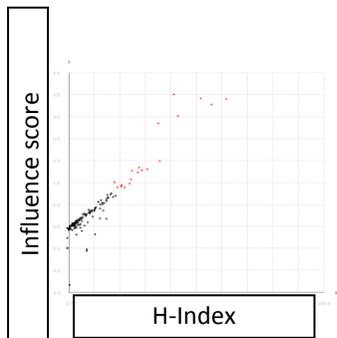

**Fig: 9 Graphical plot of Influence Score vs H-Index**

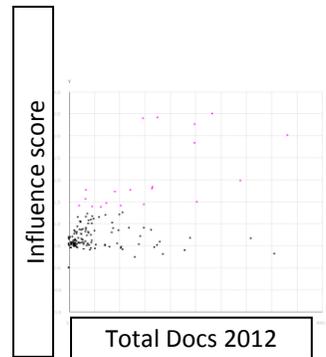

**Fig: 10 Graphical plot of Influence Score vs Total Docs 2012**

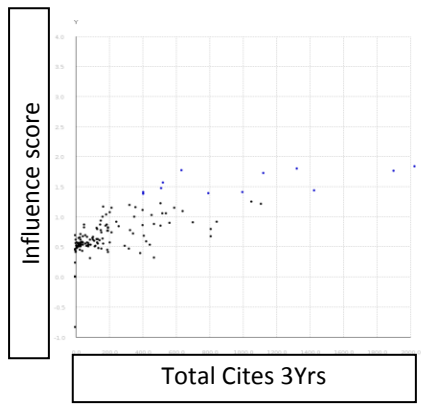

**Fig 11: Graphical plot of Influence Score vs Total Cites 3Yrs**

Looking at the graphs it is clear that H-Index explains more percentage of the scatter as shown as a result by DSRS as the scatter follows a linear pattern and also because we can say that the numbers of scatter points that lie closer to the decision boundary or the class boundary are very less. This in-turn reduces the risk involved in the class selection process while taking this factor into consideration. Hence it is also clear that H-Index defines the scatter to the Maximum extent along with the quarter in which the journal is published.

This is not the case with the other two factors here. The numbers of scatter points closer to the decision boundary are more and hence more risk is involved while making the class decision based on these factors.

## 6. Conclusion and Discussion:

In this paper, a regression analysis based method is proposed to calculate the Journal Influence Score. Comparison has been done between the rankings using SCImago Journal Rank (SJR) and the proposed method. The results show that error is minimal. The model depicts significantly high accuracy levels as each quartile match varies from 78% - 92%. This is achieved without any iterative approach and requirements of data storage.

In this model, Down-Selection based on Regression and Significance (DSRS) generates a result that shows the percentage of the variability as explained by the given dataset. Pair-wise correlation and retaining only those pairs with minimal correlation helps in reduction of input factors while maintaining reasonable accuracy. The final regression equation involves a smaller, compact set of only the principal factors. Finally a classification scheme is used for categorization of journals into "National" and "International" based on influence in order to help Libraries and repositories across the scientific and technical communities in the arduous task of categorization.

Proposed approach may be extended to a universal weight vector for all subject categories rather than evaluating separate weights for each. This might increase the percentage of errors in the model as the weights vary from categorically, but the approach would produce generic results.

Our contribution is twofold. We propose DSRS, a new metric for the evaluation of journals via the Journal Influence Score. Additionally the method proposed for computing the score is also lightweight. While various methods of computing such scores have been proposed for measuring the scientific influence of scholarly journals, typically the computation of any of these scores involves compiling the citation information pertaining to the journal under consideration. This involves significant overhead since the article citation information of not only the journal under consideration but also that of other journals for the recent few years need to be stored. Our work is motivated by the idea of developing a computationally lightweight approach that does not require any data storage, yet yields a score which is useful for measuring the influence of journals. Such a method would be especially useful to evaluate new journals, since typically there is a minimum time a journal needs to be in publication before it can be indexed. The method would of course also be applicable for listed and indexed journals available in universally accepted repositories. The screenshots provided in the appendix demonstrate this functionality amply.

The proposed model is validated using historical data from the SCImago portal. The results show that the error is small between rankings obtained using the proposed method and the SCImago Journal Rank, thus suggesting that the proposed approach is a feasible and effective alternative of calculating scientific impact of journals. Additionally clustering analysis using the features used in the computation of the influence score indicates a grouping into "National" and "International" categories. This leads to DSRIS, Down-selection and Regression based score of internationality, a discriminatory approach to classification of journals.

The SCImago approach of evaluating journal is iterative and is accomplished by using Google's page rank algorithm [8]. The problem with such approach is, if the initial rank/guesses are not a good approximation, the process has to restart. SCImago and SCOPUS require storage of significant volume of data.

DSRS does not require storage of data as the toolkit [23] demonstrates. The approach is able to estimate and forecast journal influence in a non-iterative way with very good accuracy. This is achieved by using two years of data only as opposed to processing/storing five years of data mandated by SCOPUS. Despite compromising on more than half og f the input variables from the SCiMago dataset, it is also observed that JIS method is pretty close to SJR in terms of accuracy. The following statistic bears testimony to that.

Average Difference in Journal's Ranking: ***0.12***

Median *of* difference in Journal's Ranking: ***0.08***

The regression equation, (22) as shown, has a remarkable low intercept value, implying the model does not spike/boost the default influence value of any journal, in the absence of all other input/predictor variables. These values range from 0.18 to 0.513.

Therefore, our final regression model is stable since it has predictor variables with dense data and no missing entries, unlike SCImago Journal Rank. We achieved very good accuracy by using less than half of the predictor/input variables as compared to SCImago Journal Rank. The model is applicable to journals not listed in the database as the essential parameters of DSRS are obtained by crawlers and parsers specifically written for this task. DSRS thus plays the role of a model-checker as well.

Exploring the dimensions of a journal's internationality, the authors argue internationality of a journal is dependent upon the influence a journal wields in its respective categories. However, journal influence score (JIS) as proposed and computed in present work, is a necessary but not sufficient condition for internationality measure of journal. JIS is modeled as a facilitator and complementary agent of internationality estimation. A novel metric, JIMI [19],[26] proposed by the authors' shows that JIMI and JIS are moderately correlated and therefore, unbiased and quantitative measure of "how international a journal is" should be convex combination of two. The weights to be assigned to JIMI and JIS are determined according to correlation between the two and vary by different categories of journals indexed by SCOPUS.

Reflection on the current work on JIS, the authors believe that the National and International clusters obtained by using K means clustering algorithm, are not granular enough to indicate an alignment based on the influence score. Finer levels of granularity is achieved by using JIMI and JIS together and formed the crux of ongoing and the future work by the authors in the field of internationality estimation. The outcome of the clustering technique highlights the following facts: approximately 22-30% of the SCOPUS indexed journals are not "truly influential/international" as they fail to meet the "influence threshold". The influence threshold is used as lower bound for minimal influence of a journal and is also constructed as an empirical indicator of the greatest lower bound of internationality of the journal under evaluation.

Authors do not view or judge internationality of journals as suggested in [17,18]. The objective is to score "internationality" by imparting unbiased quantitative features to it. All the five input variables are measure of such "influence" i.e. "internationality" according to author's definition. But this does not complete the definition of internationality. In order to gain a statistically unbiased and functionally correct assessment of "Reliable Measure of internationality" as more parameters are required. As a test case, if all 13 parameters from SCImago and 7 from JIMI metric [cf section 7] are included, there will be 20 features to deal with which would be a mammoth computational task. Since the authors showed already that 13 are not required and a set of only 5 predictor variables is enough, the eventual task of measuring internationality may be achieved in realizable time.

## 7. Future Work to explore Internationality

There has been a noticeable tendency among faculties to publish in so-called "International Journals" as a means of accelerating publications and enhancing the quantity of publications. On the other hand it is also hard for an author to assess which would be a suitable journal for submitting his/her work, commensurate with the quality of work. Since the tag of "internationality" is generally regarded as possessing more credibility than national journals, this is driving a volume of faculty towards publishing their work in "International Journals".

The international characteristics of scientific journals have been studied and analyzed by various researchers. Wormell, Irene [16] in their work performed a study on seven LIS journals and analyzed international characteristics of scientific journals on the basis of correlation between the geographical regions of authors, citation and subscription. They claimed that the Journal Impact Factor can be used to evaluate journals at varying time internals. Buela-Casal et al [17], in order to understand the meaning of "internationality" created a questionnaire based on eleven criteria. The questionnaire was answered by 16,056 scientists from 109 countries working in all the fields of knowledge defined by UNESCO. According to authors, publications language, online access to a publication, the editorial board of a journal and co-authors belonging to different countries are a few amongst all in order to evaluate a journal's "internationality". Authors intended to establish a definition of internationality by consensus. Gouri et al [20] developed a Web based tool similar to SCOPUS to evaluate the journals which are not indexed in SCOPUS/ISI Web of Science. The tool will provide a real time search by crawling the web and capturing the recent journals and evaluating them in terms of quality and influence.

Despite the previous and ongoing studies and research, estimating internationality of any journal remains an open problem and highly controversial. There is no known metric which clinches the issue of ranking journals according to internationality beyond reasonable doubt. There are a plethora of publishing houses in India that claim internationality of their publications, journals in particular, by registering the products (journals) with an ISSN number. The fact that an ISSN number renders "internationality" to a journal should be contested within the boundaries of intellectual propriety. A measure of "internationality" should be defined as "prestige/influence" of a journal towards scholarly dissemination of new knowledge or the promise of such dissemination amply testified via quantitative measures of that particular journal. The implosion in population of journals naturally prompts the following questions?

1. How to measure the internationality of a journal given that it is already five years in publication and possesses scientometric data and/or indexed by SCOPUS/ISI?

2. How to measure the internationality of a journal given that it is between one to five years in publication and may not possess all scientometric data and may not be indexed by SCOPUS/ISI?

3. Given a relatively new entrant (journal), how do authors trust the internationality of the journal and consider it for publication?

4. Assuming that an employee of an organization has already published in one of such journals mentioned above, how does the organization assess the publication based on the quality and internationality of the journal?

5. How to judge a publication, in short-term? If there is sufficient tolerance in terms of waiting time of three years or more, bibliometric data from different sources might be obtained. However, if the trustworthiness of a publication has to be judged in quick time, the only plausible way is to judge the source where the article is published. Is it international enough?

The present paper is a very good heuristic estimate of journal influence across Science and Technology domains, as it helps the authors to dig deeper into the "internationality" problem. The authors opine, strongly, that an international journal has very little to do with the place of publication or possession of ISSN/ISBN. Rather, it should be viewed as the realm of influence across demographic regions which in a way, depends on the scientometric parameters and impact of the journal concerned.

The current work is a prelude to a story which unfolds in various dimensions and the goal is to come up with one comprehensive measure of internationality of journals and the present paper is a first step towards that, a major one. Concurrent to the present work, the authors have developed an "internationality metric" based on the seven parameters, Total Cited Documents, International Collaboration, SNIP(Source Normalized Impact per Paper), Turnaround Time, Acceptance Ratio (Rejection Ratio), Impersonal Citation Ratio and Self Citation/Total Citation (initial assessment of internationality is done on two parameters, International Collaboration and SNIP) [19, 26]. For obvious reasons, two metrics/algorithms (the present one is statistical, the other being functional/analytical) cannot be reconciled into one paper.

Authors view Internationality as convex combination of JIS and JIMI, with appropriate weights chosen for either of these. The two metrics namely Journal Influence Score and JIMI [19, 26] complement each other and must be used in tandem to evaluate the "true" internationality and influence diffusion of Scopus indexed journals and otherwise. This is a whole body of future work and part of the problem in the long term. Convex combination of the two metrics can be represented as

**Internationality of a Journal**, $Y_I = \alpha \text{ JIS} + (1 - \alpha) \text{ JIMI}; \ 0 \leq \alpha \leq 1$

where $Y_I$ refers to the internationality score as response variable(to be sorted in decreasing order)
JIS is the internationality score obtained from metric JIS
JIMI is the score evaluated from work done using two parameters (JIMI)
and $\alpha$ is a weight deduced from the cross correlation.

JIMI considers SNIP as one of the predictor variables. Waltmann et.al. [27, 28] showed that Raw citation may be overweighed in favor of certain domains and needs to be normalized. Waltmann et. al. modified the original SNIP definition to accommodate the counter-intuitive properties of the original definition[29]. SNIP, by virtue of normalizing citations across fields, auto corrects discrepancies and counters the local influence and citation clusters. Therefore, it could be used as one of the predictor variables for internationality estimation as long as no concerted effort to manipulate the metric is observed.

For Non-Scopus journals (at least three years in publication) authors propose a "parser" to capture journals and data from Google Scholar, Scopus and Journalmetrics website. BeautifulSOUP is used to crawl the web for necessary data. This information is not entirely relevant and not self-contained either. The authors view internationality as a measure that complements journal influence. Noticeably,Scopus and SJR databases are ever-expanding. The paper, as a first cut initiative, endeavored to complete Journal Influence Score using the Scopus data as test-bed  Future work would thus include modeling internationality and indexing classes/clusters of internationality using data from Scopus/Web of Science and customized web crawler through available citation databases. DSRS, the current work is a facilitator towards that direction.

**NOTE:** All supplementary files containing figures and tables are available in a public folder [24] Revised code and tool is also available [23]

*@ as accessed on 18-09-15*

**Appendix A: DSRS: Downselection using the most significant features:**

One technique commonly used to classify workload components is by the weighted sum of their parameter values. Using $a_j$ as weight for the $j$th parameter $x_j$, the weighted sum $y$ is

$$y = \sum_{j=1}^{n} a_j x_j \qquad (12)$$

This sum can then be used to classify the components into a number of classes such as low demand or medium demand. Although this technique is commonly used in performance analysis software, in most cases, the person running the software is asked to choose the weight. Without any concrete guidelines, the person may assign weights such that workload components with very different characteristics may be grouped together, and the mean characteristics of the group may not correspond to any member.

A method of determining the weights in such situation is to find the weights $w_i$'s such that $y_j$'s provide the maximum discrimination among the components.

    A. Compute the mean and standard deviations of the variables.

$$\bar{x}_s = \frac{1}{n}\sum_{i=1}^{n} x_{si} \qquad s_{x_s}^2 = \frac{1}{n-1}\sum_{i=1}^{n}(x_{si} - \bar{x}_s)^2 \qquad (13)$$

    B. Normalize the variables to zero mean and unit standard deviation.

$$x'_s = \frac{x_s - \bar{x}_s}{S_{x_s}} \qquad (14)$$

    C. Compute the correlation among the variables:

$$R_{x_s,x_r} = \frac{(1/n)\sum_{i=1}^{n}(x_{si} - \bar{x}_s)(x_{ri} - \bar{x}_r)}{S_{x_s} S_{x_r}} \qquad (15)$$

**D.** Prepare the correlation matrix:

E. Compute the eigen values of the correlation matrix.

F. Compute the eigenvectors of the correlation matrix.

G. Obtain principal factors by multiplying the eigenvectors by the normalized vectors:

H. Compute the values of the principal factors.

I. Compute the sum and sum of squares of the principal factors. The sum must be zero. The sum of squares gives the percentage of variation explained. The process enables selection of the most influential factors for computing the influence score of journals.

Appendix B: Code for the tool is available at [23]

**Screenshots**

Fig 1. Front Screen

Fig. 2 Quarter values (1-4)

Fig 3. H-Index

Fig 4. Total Documents for 2012

**Fig 5. Total References**

**Fig 6. Cites/Docs for 2 years**

**Fig. 7 FINAL Output Screen With JIS Score**